\documentclass{article}

\usepackage{algorithm, algpseudocode}
\usepackage{amsfonts, amsmath, amsopn, amsthm, epsfig}
  \numberwithin{equation}{section}
\usepackage{graphicx, epstopdf}
\usepackage{hyperref}
\usepackage{subfigure}
\usepackage{appendix}
\usepackage{authblk}

\usepackage{calc}
\usepackage[top=2.5cm,bottom=2.3cm,inner=2.3cm,outer=2.3cm,bindingoffset=0.5cm,footskip=0.55cm]{geometry}
\theoremstyle{remark}

\newenvironment{lemma*}[2][Lemma]{\par\bgroup{\bfseries #1\ #2. }\it\ignorespaces}{\egroup}

\title{FurcaNeXt: End-to-end monaural speech separation with dynamic gated dilated temporal convolutional networks}

\author[1]{Liwen Zhang}
\author[2]{Ziqiang Shi\thanks{Corresponding author: shiziqiang@fujitsu.com; shiziqiang7@gmail.com}}
\author[1]{Jiqiang Han}
\author[3]{Anyan Shi}
\author[1]{Ding Ma}

\affil[1]{Harbin Institute of Technology, Harbin, China}
\affil[2]{Fujitsu R \& D Center, Beijing, China}
\affil[3]{Shuangfeng First, Beijing, China}

\newcommand{\BALD}{\begin{aligned}}
\newcommand{\EALD}{\end{aligned}}
\newcommand{\BALDS}{\begin{aligned*}}
\newcommand{\EALDS}{\end{aligned*}}
\newcommand{\BCAS}{\begin{cases}}
\newcommand{\ECAS}{\end{cases}}
\newcommand{\BEAS}{\begin{eqnarray*}}
\newcommand{\EEAS}{\end{eqnarray*}}
\newcommand{\BEQ}{\begin{equation}}
\newcommand{\EEQ}{\end{equation}}
\newcommand{\BIT}{\begin{itemize}}
\newcommand{\EIT}{\end{itemize}}
\newcommand{\BMAT}{\begin{bmatrix}}
\newcommand{\EMAT}{\end{bmatrix}}
\newcommand{\BNUM}{\begin{enumerate}}
\newcommand{\ENUM}{\end{enumerate}}

\newcommand{\BA}{\begin{array}}
\newcommand{\EA}{\end{array}}

\date{Feb., 2019}
\begin{document}

\maketitle

\renewcommand{\thefootnote}{\fnsymbol{footnote}}

%\footnotetext[1]{E-Mail: shiziqiang@cn.fujitsu.com}

\begin{abstract}
Deep dilated temporal convolutional networks (TCN) have been proved to be very effective in sequence modeling. In this paper we propose several improvements of TCN for end-to-end approach to monaural speech separation, which consists of 1) multi-scale dynamic weighted gated dilated convolutional pyramids network (FurcaPy), 2) gated TCN with intra-parallel convolutional components (FurcaPa), 3) weight-shared multi-scale gated TCN (FurcaSh), 4) dilated TCN with gated difference-convolutional component  (FurcaSu), that all these networks take the mixed utterance of two speakers and maps it to two separated utterances, where each utterance contains only one speaker's voice.  For the objective, we propose to train the network by directly optimizing utterance level signal-to-distortion ratio (SDR) in a permutation invariant training (PIT) style.
Our experiments on the the public WSJ0-2mix data corpus results in 18.4dB SDR improvement, which shows our proposed networks can leads to performance improvement on the speaker separation task.
\end{abstract}

%=============================================
\section{Introduction}
\label{sec:introduction}

Multi-talker monaural speech separation has a vast range of applications.
For example, a home environment or a conference environment in which many people talk, the human auditory system can easily track and follow a target speaker's voice from the multi-talker's mixed voice.
In this case, a clean speech signal of the target speaker needs to be separated from the mixed speech to complete the subsequent recognition work.
Thus it is a problem that must be solved in order to achieve satisfactory performance in speech or speaker recognition tasks. There are two difficulties in this problem, the first is that since we don't have any priori information of the user, a truly practical system must be speaker-independent. The second difficulty is that there is no way to use the beamforming algorithm for a single microphone signal. Many traditional methods, such as computational auditory scene analysis (CASA)~\cite{wang2006computational,shao2006model,hu2013unsupervised}, Non-negative matrix factorization (NMF)~\cite{smaragdis2007convolutive, le2015sparse}, and probabilistic models~\cite{virtanen2006speech}, do not solve these two difficulties well.

More recently, a large number of techniques based on deep learning are proposed for this task. These methods can be briefly grouped into three categories. The first category is based on deep clustering (DPCL)~\cite{hershey2016deep,isik2016single}, which maps the time-frequency (TF) points of the spectrogram into the embedding vectors, then these embedding vectors are clustered into several classes corresponding to different speakers, and finally these clusters are used as masks to inversely transform the spectrogram to the separated clean voices; the second is the permutation invariant training (PIT)~\cite{kolbaek2017multitalker,yu2017permutation}, which solves the label permutation problem by minimizing the lowest error output among all possible permutations for $N$ mixing sources assignment; the third category is end-to-end speech separation in time-domain~\cite{luo2017tasnet,luo2018tasnet,venkataramani2017adaptive}, which is a natural way to overcome the obstacles of the upper bound source-to-distortion ratio improvement (SDRi) in short-time Fourier transform (STFT) mask estimation based methods and real-time processing requirements in actual use.

This paper is based on the end-to-end method~\cite{luo2017tasnet,luo2018tasnet,venkataramani2017adaptive}, which has achieved better results than DPCL based or PIT based approaches. Since most DPCL and PIT based methods use STFT as front-end. Specifically, the mixed speech signal is first transformed from one-dimensional signal in time domain to two-dimensional spectrum signal in TF domain, and then the mixed spectrum is separated to result in spectrums corresponding to different source speeches by a deep clustering or mask estimation method, and finally the cleaned source speech signal can be restored by an inverse STFT on each spectrum. This framework has several limitations. Firstly, it is unclear whether the STFT  is the optimal (even assume the parameters it depends on are optimal, such as size and overlap of audio frames, window type and so on) transformation of the signal for speech separation. Secondly, most STFT based methods often assumed that the phase of the separated signal to be equal to the mixture phase, which is generally incorrect and imposes an obvious upper bound on separation performance by using the ideal masks. As an approach to overcome the above problems, several
speech separation models were recently proposed that operate
directly on time-domain speech signals~\cite{luo2017tasnet,luo2018tasnet,venkataramani2017adaptive}.
Inspired by these first results, we propose FurcaNeXt, which is a general name for a series of fully end-to-end time-domain separation methods, includes 1) multi-scale dynamic weighted gated dilated convolutional pyramids network (FurcaPy): due to the influence of different word lengths or different speech speeds, multiple branches of a variety of temporal receipt field scales are introduce to characterize speech, and the weights of different scales are automatically determined by a ``weightor'' network;
2) deep gated dilated temporal convolutional networks (TCN) with intra-parallel convolutional components (FurcaPa): replace
two convolutional related modules in each dilated convolutional module by two intra-parallel convolutional modules
, which can reduce the variance of this model. The intra-parallel convolutional modules replicate weight matrices and take the average from the feature maps produced by those layers. This convenient technique can effectively improve separation performance.
3) weight-shared multi-scale gated TCN (FurcaSh): a simple design is proposed to achieve the functions of FurcaPy  but without increasing the number of network parameters.
4) dilated TCN with gated difference-convolutional component (FurcaSu): inspired by the work of Highway network~\cite{srivastava2015highway}, in which two additional  non-linear  transformations acts as gates that can dynamically
pass part of its inputs and suppress the other part,
conditioned on the input itself. Authors simplify the Highway network through multiple ways~\cite{yousef2018accurate}. After further simplification we propose to use two identical transformation function branches to implemented a simplified version of the highway network module.

The remainder of this paper is organized as follows:  section 2 introduces monaural
speech separation with TCN. Section 3 describe our proposed FurcaNeXt and the separation algorithm in detail. The experimental
setup and results are presented in Section 4. We conclude this
paper in Section 5.

\section{Speech separation with TCN}
\label{sec:tcn}

In this section, we review the formal definition of the monaural speech separation task and the original TCN architecture.

The goal of monaural speech separation is to estimate the individual target signals from a linearly mixed single-microphone signal, in which the target signals overlap in the TF domain.
Let $x_i(t),i=1,..,S$ denote the $S$ target speech signals and  $y(t)$ denotes the
mixed speech respectively. If we assume the target signals are linearly mixed, which can be represented as:
\begin{equation*}
y(t)=\sum_{i=1}^{S}x_i(t),
\end{equation*}
then monaural speech separation aims at estimating individual target signals from
given mixed speech $y(t)$. In this work it is assumed that the number of target signals is known.

\begin{figure}%[htp]
%\vspace{-0.4in}
\centering
%\begin{center}
\hspace{-5mm}
\includegraphics[width=0.7\linewidth]{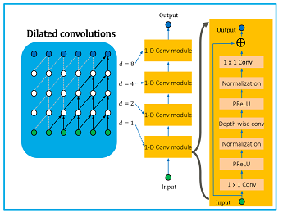}
\hspace{-5mm}
%\end{center}
%\vspace{-2.5mm}
\caption{
The structure of TCN.
}
\label{tcn}
%\vspace{-2.5mm}
\end{figure}

In order to deal with this ill-posed problem, Luo et al.~\cite{luo2018tasnet} introduce TCN~\cite{lea2016temporal,bai2018empirical} to do this task. TCN is proposed as an alternative to RNN in various tasks~\cite{lea2016temporal,bai2018empirical}.  Each layer in the TCN contains a 1-D convolution block with an increased dilation factor. The dilation factor is increased exponentially to ensure a suitable large temporal context window to take advantage of the long range dependence of the speech signal, as shown in Figure~\ref{tcn}. Dilated convolution has made a huge success in WaveNet for audio generation~\cite{van2016wavenet}. Dilated convolutions with different dilations have different receptive fields.  Stacked dilated convolution  provides a very large receptive fields for the network with only a few layers, because the dilation range grows exponentially. This allows the network to capture the temporal dependence of various resolutions with  the input sequences. The TCN introduces a time hierarchy: the upper layer can access longer input subsequences and learn representations on larger time scales. Local information from lower layers spreads through the hierarchy by means of residuals and skip connections.

There are two important elements in the original TCN~\cite{bai2018empirical} as shown in Figure~\ref{tcn}, one is the dilated convolutions, and the other is residual connections. Dilated convolutions follow the work of~\cite{van2016wavenet}, it is defined as
\begin{equation*}
(x\ast_d k)(p)=\sum_{s+dt=p}x(s)k(t),
\end{equation*}
where $x$ is the 1-D input signal, $k$ is the filter(aka kernel), and $d$ is the dilation factor. Therefore, dilation is equivalent to introducing a fixed step size between every two adjacent filter taps. The general way to increase the receipt field of the TCN is to increase the dilation
factor $d$. In this work we increase $d$ exponentially with the depth of the network and $d=2$ as shown in Figure~\ref{tcn}, and this TCN has four layers of 1-D Conv modules with dilation factors of $1, 2, 4, 4$ respectively. As shown in Figure~\ref{tcn}, Each 1-D Conv module is a residual block~\cite{he2016deep}, which contains one layer of dilated convolution (Depth wise conv~\cite{howard2017mobilenets}), two layers of 1$\times$ 1 convolutions (1$\times$ 1 Conv), two non-linearity activation layers (Parametric Rectified Linear Unit, PReLU~\cite{he2015delving}), and two normalization layers (Normalization). For normalization, we applied global normalization~\cite{luo2018tasnet} to the convolutional filters.

\begin{figure}%[htp]
%\vspace{-0.4in}
\centering
%\begin{center}
\hspace{-5mm}
\includegraphics[width=0.7\linewidth]{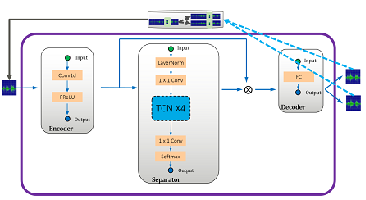}
\hspace{-5mm}
%\end{center}
%\vspace{-2.5mm}
\caption{
The pipeline of TCN based speech separation in~\cite{luo2018tasnet}.
}
\label{tcnline}
%\vspace{-2.5mm}
\end{figure}

Luo et al. proposed a TCN based speech separation method~\cite{luo2018tasnet}, which consists of three processing stages, as shown in Figure~\ref{tcnline}: encoder (Conv1d is followed by a PReLU), separator (consisted in the order by a LayerNorm, a 1$\times$1conv, 4 TCN layers, 1$\times$1conv, and a softmax operation) and decoder (a FC layer).  First, the encoder module is used to convert short segments of the mixed waveform into their corresponding representations. Then, the representation is used to estimate the multiplication function (mask) of each source and each encoder output for each time step. The source waveform is then reconstructed by transforming the masked encoder features using a linear decoder module.

\section{Speech separation with FurcaNeXt}
\label{sec:furcanext}

\begin{figure}%[htp]
%\vspace{-0.4in}
\centering
%\begin{center}
\hspace{-5mm}
\includegraphics[width=0.7\linewidth]{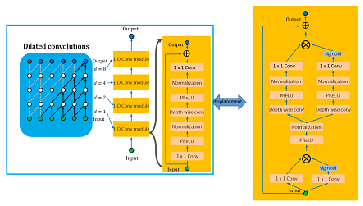}
\hspace{-5mm}
%\end{center}
%\vspace{-2.5mm}
\caption{
The structure of gated TCN.
}
\label{gated_tcn}
%\vspace{-2.5mm}
\end{figure}

The main work of this paper is to make several improvements to the TCN (Figure~\ref{tcn}) and TCN based framework (Figure~\ref{tcnline}) for speech separation. First, we introduced gating operations in this TCN, as shown in Figure~\ref{gated_tcn}. Nonlinear gated activation had been used in prior work on sequence modeling~\cite{van2016wavenet,dauphin2016language}, it can control the flow of information and  may help the network to model more complex interactions. Two gates are added to each 1-D convolutional module in the plain TCN, one is corresponding to the first 1$\times$1 convolutional layer in the 1-D convolutional module, the other is corresponding to all the layers from the depth-wise convolutional layer to the output 1$\times$1 convolutional layer. This gated TCN based speech separation pipeline is called FurcaPorta in this work.

\subsection{FurcaPy: Multi-scale dynamic weighted gated dilated convolutional pyramids network}

%Since in real life the utterance always have the feature  of  temporal scale variation caused by different word lengths and  pronunciation characteristics (e.g. speed) of different people, thus different temporal receipt fields may help in speech separation. The temporal receipt field is fixed in previous network structure. In order to remedy the temporal scale variation problem, a multi-scale dynamic weighted \underline{\textbf{py}}ramids gated TCNs based pipeline which is called FurcaPy is proposed as shown in Figure~\ref{furcapy} and there are three kinds of different temporal receipt fields in this description. Further more a ``weightor'' module is designed to determine which   temporal receipt field is more suitable for current input utterance, that means the weights of different gated TCNs are determined dynamically by a ``weightor'' network for each utterance. The ``weightor'' is composed of a common multi layer 1-D convolutional neural network as shown in Figure~\ref{furcapy}.

\begin{figure}%[htp]
%\vspace{-0.4in}
\centering
%\begin{center}
\hspace{-5mm}
\includegraphics[width=0.7\linewidth]{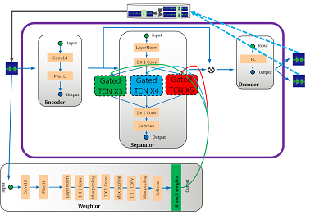}
\hspace{-5mm}
%\end{center}
%\vspace{-2.5mm}
\caption{
The structure of FurcaPy.
}
\label{furcapy}
%\vspace{-2.5mm}
\end{figure}

Since in real life the utterance always have the feature of  temporal scale variation caused by different word lengths and  pronunciation characteristics (e.g. speed) of different people, thus different temporal receipt fields may help in speech separation. The temporal receipt field is fixed in previous network structure. In order to remedy the temporal scale variation problem, a multi-scale dynamic weighted \underline{\textbf{py}}ramids gated TCNs based pipeline which is called FurcaPy is proposed as shown in Figure~\ref{furcapy} and there are three kinds of different temporal receipt fields in this description. FurcaPy's encoder and decoder are the same as the previous FurcaPorta, they differ only in the separator. In the separator of FurcaPy, each branch in the pyramid consists of a different number of  gated TCNs. The length of the temporal receptive field of each branch is several times the length of the temporal receiving field of a single gated TCN.
If the receptive field of a single gated TCN is assumed to be $L$, then the length of the receptive field of all branches in the Figure~\ref{furcapy} is 3$L$,4$L$, and 5$L$ respectively.
The total output is obtained by weighted averaging the outputs of the different branches corresponding to different receipt fields. Additionally, a ``weightor'' module is designed to determine which  temporal receipt field is more suitable for current input utterance signal, that means the weights of different gated TCNs are determined dynamically by a ``weightor'' network for each utterance. The ``weightor'' is composed of a common multi layer 1-D convolutional neural network as shown in Figure~\ref{furcapy} and it consist of Conv1d, PReLU, LayerNormal, 3 layers of 1$\times$1 Conv and max pooling, and Softmax.

\subsection{FurcaSh: Weight-shared multi-scale gated TCN}

%The parameters of FurcaPy are basically twice the FurcaPorta. In fact, there is another way without additional parameters to do multi-scale temporal receipt fields, which can be achieved by average through outputs of different gated TCN layers and average through outputs of different layers corresponding to different dilate factors in each 1-D convolutional module as shown in Figure~\ref{furcash} and Figure~\ref{furcasha} respectively. Thus the weights of different temporal receipt fields are \underline{\textbf{sh}}ared. This structure is called FurcaSh.

\begin{figure}%[htp]
%\vspace{-0.4in}
\centering
%\begin{center}
\hspace{-5mm}
\includegraphics[width=0.7\linewidth]{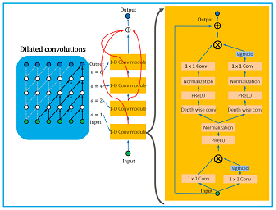}
\hspace{-5mm}
%\end{center}
%\vspace{-2.5mm}
\caption{
The structure of FurcaSh.
}
\label{furcash}
%\vspace{-2.5mm}
\end{figure}

\begin{figure}%[htp]
%\vspace{-0.4in}
\centering
%\begin{center}
\hspace{-5mm}
\includegraphics[width=0.7\linewidth]{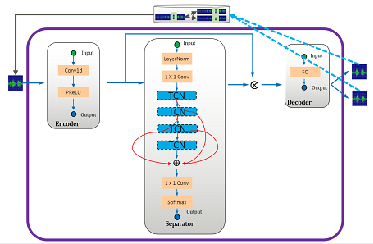}
\hspace{-5mm}
%\end{center}
%\vspace{-2.5mm}
\caption{
The structure of FurcaSh.
}
\label{furcasha}
%\vspace{-2.5mm}
\end{figure}

FurcaPy will increase the number of parameters of the network several times, and the processing speed of the network will decrease a lot. In many cases, there is no way to meet the requirements of real-time processing for such network.
In order to deal with these problems, a new structure is proposed that can achieve the two-level multi-scale receptive fields without increasing the number of network parameters. As shown in Figure~\ref{furcash} and  Figure~\ref{furcasha}, two levels of multi-scale temporal receptive fields is introduced, one is in the dilated 1-D conv module level, that is the outputs corresponding to different dilated factors are summed and averaged to result in the final output of this gated TCN; the other is that since there are 4 gated TCNs in the FurcaSh pipeline, the outputs of different gated TCNs are summed and averaged to result in the separator.
So there are two different levels of  multi-scale temporal receipt field in this structure.

\subsection{FurcaPa: Deep gated dilated temporal convolutional networks (TCN) with intra-parallel convolutional components}

%The performance of a single predictive model can always be improved by ensemble, that is to combines a set of independently trained networks.
%The most commonly used method is to do the average of the model, which can at least  help to reduce the variance of the performance. As shown in Figure~\ref{furcapa}, in the different layers of each 1-D convolutional module of gated TCN in FurcaPorta, two identical \underline{\textbf{pa}}rallel branches are added. This structure is called FurcaPa.

\begin{figure}%[htp]
%\vspace{-0.4in}
\centering
%\begin{center}
\hspace{-5mm}
\includegraphics[width=0.9\linewidth]{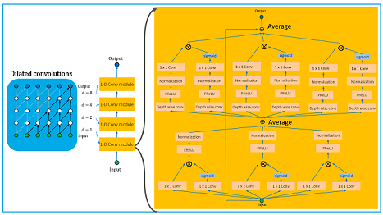}
\hspace{-5mm}
%\end{center}
%\vspace{-2.5mm}
\caption{
The structure of FurcaPa.
}
\label{furcapa}
%\vspace{-2.5mm}
\end{figure}

The performance of a single predictive model can always be improved by ensemble, that is to combines a set of independently trained networks.
The most commonly used method is to do the average of the model, which can at least  help to reduce the variance of the performance. As shown in Figure~\ref{furcapa}, in the different layers of each 1-D convolutional module of gated TCN in FurcaPorta, two identical \underline{\textbf{pa}}rallel branches are added. This structure is called FurcaPa.
The total output of each intra-parallel convolutional components is obtained by averaging the outputs of all the different branches. In each single dilated 1-D convolutional module layer, two intra-parallel convolutional components are introduced, the first one is near the input layer and covers the Conv1d, PReLU, and Normalization layers; the other one is near the output and it covers the rest all layers, including the Depthwise conv, PReLu, Normalization and 1$\times$1 Conv layers. The reason why we do this ensemble in two places is to reduce the sub-variances of each block.

\subsection{FurcaSu: Dilated TCN with gated difference-convolutional component}

%The operation of element-wise subtraction may increase the complexity of the model and result in better performance. A difference(\underline{\textbf{su}}btraction)-convolutional component is added to the gated TCN in the FurcaPorta as shown in Figure~\ref{furcasu}.

\begin{figure}%[htp]
%\vspace{-0.4in}
\centering
%\begin{center}
\hspace{-5mm}
\includegraphics[width=0.7\linewidth]{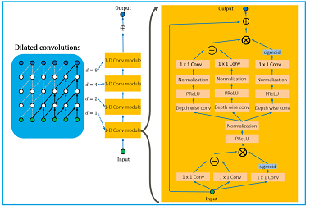}
\hspace{-5mm}
%\end{center}
%\vspace{-2.5mm}
\caption{
The structure of FurcaSu.
}
\label{furcasu}
%\vspace{-2.5mm}
\end{figure}

Highway network can be simplified and generalized to have better performance~\cite{yousef2018accurate}. Follow the work of~\cite{yousef2018accurate}, we further simplify the Highway network module, as shown in Figure~\ref{furcasu}, in each single dilated 1-D convolutional module layer, two Highway network module or gated difference-convolutional components as we called are introduced, the first one is near the input layer and covers the Conv1d, PReLU, and Normalization layers; the other one is near the output and it covers the rest all layers, including the Depthwise conv, PReLu, Normalization and 1$\times$1 Conv layers.
Different from the original use of three different transformation functions, in order to simplify the design and improve performance, here we use three identical transformation branches, one branch as the attention gates, the other two are  signal transformations, and their results are subtracted and then gated.

\subsection{Perceptual metric: Utterance-level SDR objective}
\label{sec:loss}

Since the loss function of many STFT-based methods is not directly applicable to waveform-based end-to-end speech separation, perceptual metric based loss function is tried in this work. The perception of speech is greatly affected by distortion~\cite{yang1998performance,assmann2004perception}.
Generally in order to evaluate the performance of speech separation, the BSS\_Eval metrics signal-to-distortion ratio (SDR), signal-to-Interference
ratio (SIR), signal-to-artifact ratio (SAR)~\cite{fevotte2005bss,vincent2006performance},
and short-time objective intelligibility (STOI)~\cite{taal2010short}
have been often employed. In this work we directly use SDR, which is most commonly used metrics to evaluate the performance of source separation, as the training objective. SDR measures the amount of distortion introduced by the output signal and define it as the ratio between the energy of the clean signal and the energy of the distortion.

SDR captures the overall separation quality of the algorithm. There is a subtle problem here. We first concatenate the outputs of FurcaNet into a complete utterance and then compare with the input full utterance to calculate the SDR in the utterance level instead of calculating the SDR for one frame at a time. These two methods are very different in ways and performance. If we denote the output of the network by $s$, which should ideally be equal to the target source $x$, then SDR can be given as~\cite{fevotte2005bss,vincent2006performance}
\begin{eqnarray*}
 \tilde{x}&=&\frac{\langle x , s \rangle}{\langle x , x \rangle} x, \\ e&=&\tilde{x}-s,\\ \text{SDR} &=& 10*\text{log}_{10}\frac{\langle \tilde{x} , \tilde{x} \rangle}{\langle e , e \rangle}.
\end{eqnarray*}
Then our target is to maximize SDR or minimize the negative SDR as loss function respect to the $s$.

In order to solve tracing and permutation problem, the PIT training criteria~\cite{kolbaek2017multitalker,yu2017permutation} is employed in this work. We calculate the SDRs for all the permutations, pick the maximum one, and take the negative as the loss. It is called the uSDR loss in this work.

\section{Experiments}
\label{sec:experiments}

\subsection{Dataset and neural network}
\label{ssec:dataset}

We evaluated our system on two-speaker speech separation problem using WSJ0-2mix dataset~\cite{hershey2016deep,isik2016single}, which contains 30 hours of training and 10 hours of validation data. The mixtures are generated by randomly selecting 49 male and 51 female speakers and utterances in Wall Street Journal (WSJ0) training set si\_tr\_s, and mixing them at various signal-to-noise ratios (SNR) uniformly between 0 dB and 5 dB. 5 hours of evaluation set is generated in the same way, using utterances from 16 unseen speakers from si\_dt\_05 and si\_et\_05 in the WSJ0 dataset.

We evaluate the systems with the SDR improvement (SDRi)~\cite{fevotte2005bss,vincent2006performance} metrics used in~\cite{isik2016single,luo2018speaker,xu2018single,chen2017deep,kolbaek2017multitalker}.
The original SDR, that is the average SDR of mixed speech $y(t)$ for the original target speech $x_1(t)$ and $x_2(t)$ is 0.15.
Table~\ref{tab:sdri} lists the average SDRi obtained by the different structures in FurcaNeXt and almost all the results in the past two years, where IRM means the ideal ratio mask
\begin{equation}
M_s=\frac{|X_s(t,f)|}{\sum_{s=1}^{S}|X_s(t,f)|}
\label{eq:irm}
\end{equation}
applied to the STFT $Y(t,f)$ of $y(t)$ to obtain the separated speech, which is evaluated to show the upper bounds of STFT based methods, where $X_s(t,f)$ is the STFT of $x_s(t)$.

In this experiment, as baselines, we reimplemented several classical approaches, such as DPCL~\cite{hershey2016deep}, TasNet~\cite{luo2017tasnet} and Conv-TasNet~\cite{luo2018tasnet}. Table~\ref{tab:sdri} lists the SDRis obtained by our methods and almost all the results in the past two years, where IRM means the ideal ratio mask.
Compared with these baselines an average increase of nearly 2.6dB SDRi is obtained. FurcaPy has achieved the most significant performance improvement compared with baseline systems, and it break through the upper bound of STFT based methods a lot (nearly 6dB).

\begin{table}[th]
\caption[sdri]{SDRi (dB) in a comparative study of different separation methods on the WSJ0-2mix dataset. * indicates our reimplementation of the corresponding method.}\label{tab:sdri}
\centering
\begin{tabular}{|c|c|}
\hline
Method & SDRi  \\
\hline
DPCL~\cite{hershey2016deep} & 5.9  \\
\hline
uPIT-BLSTM~\cite{yu2017permutation} & 10.0 \\
\hline
cuPIT-Grid-RD~\cite{xu2018single} & 10.2 \\
\hline
DANet~\cite{chen2017deep} & 10.5 \\
\hline
ADANet~\cite{luo2018speaker} & 10.5 \\
\hline
DPCL* & 10.7  \\
\hline
DPCL++~\cite{isik2016single} & 10.8  \\
\hline
CBLDNN-GAT~\cite{li2018cbldnn} & 11.0 \\
\hline
TasNet~\cite{luo2017tasnet} & 11.2 \\
\hline
TasNet* & 11.8 \\
\hline
Chimera++~\cite{wang2018alternative} & 12.0 \\
\hline
FurcaX~\cite{shi2019furcax} & 12.5 \\
\hline
IRM & 12.7 \\
\hline
FurcaNet~\cite{shi2019furcanet} & 13.3 \\
\hline
Conv-TasNet~\cite{luo2018tasnet} & 15.0 \\
\hline
Conv-TasNet* & 15.8 \\
\hline
FurcaPorto & 17.3 \\
\hline
FurcaSu & 17.9 \\
\hline
FurcaSh & 18.0 \\
\hline
FurcaPa & 18.2 \\
\hline
FurcaPy & 18.4 \\
\hline
\end{tabular}
\end{table}

\section{Conclusion}

In this paper we investigated the effectiveness of deep dilated temporal convolutional networks modeling for multi-talker monaural speech separation. 
We propose a series structure under the name of FurcaNeXt do to speech separation.
Benefits from the strength of end-to-end processing, the  novel gating mancinism and dynamic improvements, the best performance of structure in FurcaNeXt achieve 18.4dB SDRi on the the public WSJ0-2mix data corpus, results in 16\% relative improvement, and we achieve the new state-of-the-art on the public WSJ0-2mix data corpus.
For further work, although SDR is widely used and can be useful, but it has some weaknesses~\cite{roux2018sdr}. In the future, maybe we can use SNR to evaluation our models. It would be interesting to see how consistent the SDR and SNR are.

\section{Acknowledgment}
We would like to thank Jian Wu at Northwestern Polytechnical University, Yi Luo at Columbia University, and Zhong-Qiu Wang at Ohio State University for valuable discussions on WSJ0-2mix database, DPCL, and end-to-end speech separation.

% ---- Bibliography ----
%\bibliographystyle{abbrv}  %this one
%\bibliography{IEEEabrv,CommunityBIB-Jerry.bib}
\bibliographystyle{splncs03}
\bibliography{furcanext_arxiv}

\end{document}